\documentclass[%
preprint,
 superscriptaddress,
 showpacs,
 amsmath,amssymb,
 aps,
prc, 
]{revtex4-1}

\usepackage{graphicx}
\usepackage{dcolumn}
\usepackage{bm}

\begin{document}


\title{
Test of the Conserved Vector Current Hypothesis
by $\beta$-ray Angular Distribution Measurement in the Mass-8 System
 }

\author{T.~Sumikama}
\affiliation{Department of Physics, Osaka University, 1-1 Machikaneyama, Toyonaka, Osaka 560-0043, Japan}
\affiliation{Department of Physics, 
Faculty of Science and Technology, 
Tokyo University of Science, Noda, Chiba 278-8510, Japan}
\author{K.~Matsuta}
\affiliation{Department of Physics, Osaka University, 1-1 Machikaneyama, Toyonaka, Osaka 560-0043, Japan}
\author{T.~Nagatomo}
\affiliation{Department of Chemistry, International Christian University, Mitaka, Tokyo 181-8585, Japan}
\author{M.~Ogura}
\author{T.~Iwakoshi}
\author{Y.~Nakashima}
\author{H.~Fujiwara}
\author{M.~Fukuda}
\author{M.~Mihara}
\affiliation{Department of Physics, Osaka University, 1-1 Machikaneyama, Toyonaka, Osaka 560-0043, Japan}
\author{K.~Minamisono}
\affiliation{National Superconducting Cyclotron Laboratory, 
Michigan State University, East Lansing, Michigan 48824, USA}
\author{T.~Yamaguchi}
\affiliation{Department of Physics, Saitama University, Saitama 338-8570, Japan}
\author{T.~Minamisono}
\affiliation{Fukui University of Technology, 3-6-1 Gakuen, Fukui 910-8505, Japan}

\date{\today}

\begin{abstract}

The $\beta$-ray angular correlations for the spin alignments
of $^8$Li and $^8$B have been observed 
in order to test the conserved vector current (CVC) hypothesis.
The alignment correlation terms were combined with 
the known $\beta$-$\alpha$-angular correlation terms 
to determine all the matrix elements
contributing to the correlation terms. 
The weak magnetism term, $7.5\pm0.2$, 
deduced from the $\beta$-ray 
correlation terms was consistent with the CVC prediction
$7.3\pm0.2$, 
deduced from the analog-$\gamma$-decay measurement  based on the CVC hypothesis.
However, there was no consistent CVC prediction
for the second-forbidden term associated with the weak vector current.  
The experimental value for the second-forbidden term was $1.0\pm0.3$, 
while the CVC prediction was  $0.1\pm0.4$ or $2.1\pm0.5$. 
\end{abstract}


\maketitle

\section{Introduction}
In the standard electroweak model, 
the weak vector current of nucleons and 
the isovector part of the electromagnetic current 
form a single isovector electroweak current \cite{Towner95}. 
The conserved vector current (CVC) hypothesis
is analogous to the electromagnetic current conservation law. 
The weak vector current is 
conserved despite the influence of strong interactions 
such as the contribution from the pion cloud around a nucleon. 
The minimum conditions required for the CVC hypothesis
are the universality of the vector coupling constant $g_{\rm V}$ and
the absence of the induced scalar term $g_{\rm S}$ in the weak nucleon current. 
The universality of $g_{\rm V}$ is confirmed at the level of $1.2 \times 10^{-4}$
and $g_{\rm S}$ is limited to $m_e g_{\rm S}/2M_ng_{\rm V} = -(0.0011\pm 0.0013)$ 
from 20 superallowed $0^+\rightarrow 0^+$ $\beta$ decays \cite{Hardy09}, 
where $M_n$ and $m_e$ are the nucleon and electron masses, respectively.
In this article, the coupling constant and 
the induced term in the weak nucleon current, 
including $g_{\rm V}$ and $g_{\rm S}$,  
are written following Holstein's expression \cite{Holstein74}.  

In addition, the so-called strong CVC hypothesis demands that 
the weak vector current is paired with the isovector electromagnetic current. 
For the isospin triplet state, 
the strong CVC requires that 
a matrix element from the weak vector current for $\beta$ decay 
is identical to one from the isovector electromagnetic current for analog-$\gamma$ decay.  
To test the strong CVC hypothesis, 
the weak magnetism term $a_{\rm WM}^\beta$ 
has been compared with the CVC prediction 
deduced from the isovector M1 component of 
the analog-$\gamma$ transition strength 
\cite{Grenacs85}.  
The term $a_{\rm WM}^\beta$ was detected 
using one of the following: a spectral shape factor, 
a $\beta$-ray angular correlation 
with a spin orientation, or 
a correlation with a delayed $\alpha$ or $\gamma$ ray 
\cite{Lee63, Grenacs85,Minamisono02-2,DeBraeckeleer95}. 
The experimentally determined $a_{\rm WM}^\beta$ may include
a possible $G$-parity irregular term, $g_{\rm II}$, 
in the weak axial-vector current; 
therefore, the strong CVC has been tested 
under the assumption of $G$-parity conservation. 
Among those studies, Minamisono {\it et al.~}\cite{Minamisono02-2}  determined 
the most accurate $a_{\rm WM}^\beta$ in the mass $A=12$ system.  
In their paper, the $g_{\rm II}$ was determined 
using the CVC prediction, $a_{\rm WM}^{\rm CVC}$, for the weak magnetism term. 
When $G$-parity conservation was assumed,
the strong CVC was confirmed 
as $a_{\rm WM}^\beta/a_{\rm WM}^{\rm CVC}=1.04\pm0.03$ \cite{Minamisono11}.

Earlier, 
the strong CVC in the $A=8$ system was tested 
using the $\beta$-$\alpha$ angular correlation terms of $^8$Li and $^8$B
\cite{8System, Tribble75,McKeown80}.
Among the previous measurements, 
those by Tribble and McKeown \cite{Tribble75, McKeown80} were performed 
for a wide energy range of $\beta$ rays, and 
the mirror difference $\delta^-_{\beta\alpha}$ of the $\beta$-$\alpha$ angular correlation term 
was determined. 
$\delta^-_{\beta\alpha}$ has a contribution of $a_{\rm WM}^\beta$ and 
a second-forbidden term $a_{\rm WE2}^\beta$ associated with the weak vector current.  
While the Tribble's data did not reproduce 
the kinematic shift term for the angular correlation, 
the McKeown's data reproduced this term properly,  and 
the result was consistent with the CVC prediction value
$\delta^-_{\rm CVC}$ \cite{DeBraeckeleer95} 
as $\delta^-_{\beta\alpha}/\delta^-_{\rm CVC}=0.93\pm0.03\pm0.05$,
where the first uncertainty was from the $\beta$-$\alpha$ measurement and 
the second one was from the CVC prediction. 

The $\beta$-ray angular distributions of $^8$Li and $^8$B are given 
by a combination of several matrix elements; not only $a_{\rm WM}^\beta$ and 
$a_{\rm WE2}^\beta$ but also the Gamow-Teller, 
 axial charge, and second forbidden terms 
from the axial-vector currents. 
In spite of this complexity, 
we previously showed \cite{Sumikama08-1} that 
$a_{\rm WM}^\beta$ and $a_{\rm WE2}^\beta$
could be determined separately by combining the alignment correlation term and 
the $\beta$-$\alpha$ angular correlation term.
Thus, the strong CVC can be tested for the second-forbidden transition for the first time.

In our previous letter \cite{Sumikama08-1}, 
we reported the measurement of the $\beta$-ray angular correlation term 
from the spin aligned $^8$Li and $^8$B ($J^\pi = 2^+$) and 
the limitation of $g_{\rm II}$
under the assumption of CVC. 
In the present study, we reanalyzed the data with the assumption 
of $G$-parity conservation in order to test the strong CVC hypothesis
for the weak magnetism and for the second-forbidden transition separately.

\section{Beta-ray angular correlation terms and analog Gamma decay}

The two kinds of $\beta$-ray angular correlation term, 
i.e., the alignment correlation term and 
the $\beta$-$\alpha$ angular correlation term, 
are similar to each other.
The alignment correlation term is associated with
the spin alignment of parent nucleus.
Because $\beta$-delayed $\alpha$ particles 
are emitted in the direction perpendicular to 
the angular momentum of the daughter nucleus $^8$Be ($J^\pi=2^+$),
the $\beta$-$\alpha$ angular correlation term is 
associated with the spin alignment of the daughter nucleus. 
As a result, the alignment correlation term and 
the $\beta$-$\alpha$ angular correlation term
have the same formula except for the signs 
of several second-forbidden terms. 
This complementary relationship allows  
all the matrix elements to be separately determined, as follows.

The $\beta$-ray angular distribution from purely spin-aligned nuclei
is given by
$W(E,\theta_{I\beta})\propto 
pE(E_0-E)^2 \{B_0(E)
    + {\cal A} B_2(E)P_2(\cos\theta_{I\beta})\}$,
where $p$, $E$, $E_0$, and $\theta_{I\beta}$ are 
the $\beta$-ray momentum, energy, end-point energy, 
and ejection angle with respect to the spin-orientation axis, respectively.
$P_2$ is the Legendre polynomial.
The $^8$Li and $^8$B nuclei decay to the broad first excited state of $^8$Be,
thus the end-point energy $E_0$ is given as $E_0=E_{\rm max}-E_x$. 
$E_{\rm max}$ is the energy release during the $\beta$ decay 
to the $^8$Be ground state, 
while $E_x$ is the excitation energy of $^8$Be. 
The nuclear-spin alignment ${\cal A}=(2a_{+2}-a_{+1}-2a_0-a_{-1}+2a_{-2})/2$
is given by the population $a_m$ of the magnetic substate $m$, 
with $\sum a_m=1$.
The alignment correlation terms $B_2(E)/B_0(E)$
for $^8$Li and $^8$B 
are given by $K(E,0)$ 
in \cite{Holstein74} as
\begin{eqnarray}
K(E,s)&=&
  -\frac{E}{3M_n}
         \left[\frac{1}{A}\pm \frac{b}{Ac}-
           \frac{d_{\rm I}}{Ac}\mp \frac{g_{\rm II}}{g_{\rm A}}\right.\nonumber\\
&&+\frac{(-)^s}{\sqrt{14}}\left\{\pm \frac{f}{Ac}\frac{E_0+2E}{E_0}
+\frac{3}{2}\frac{j_2}{A^2c}\frac{E_0-2E}{M_n}\right\}
\nonumber\\
&&\left.- \frac{3}{\sqrt{35}}\frac{j_3}{A^2c}\frac{E}{M_n}\right],
\label{Eq:B2B0}
\end{eqnarray}
where $g_{\rm A}$ is the axial-vector coupling constant, 
$c$ is the Gamow-Teller matrix element, $b$ is the weak magnetism matrix element, 
$d_{\rm I}$ is the axial charge, 
$f$ is the second-forbidden term from the vector current,
$j_2$ and $j_3$ are the second-forbidden terms 
from the axial-vector current, 
and $A$ is the mass number of the nucleus. 
$a_{\rm WM}$ and $a_{\rm WE2}$ are given by the ratios
$a_{\rm WM}=b/Ac$ and $a_{\rm WE2}=f/Ac$. 
The $\beta$-$\alpha$ angular correlation term, on the other hand,  
is given by $W(E,\theta_{\beta\alpha})\propto pE(E_0-E)^2 
\{ 1+a_\mp (E)\cos\theta_{\beta\alpha} +p_\mp (E)\cos^2\theta_{\beta\alpha}\}$,
where $\theta_{\beta\alpha}$ is the angle between the momenta of $\beta$ and $\alpha$ rays.
$a_\mp(E)$ is the kinematic shift term associated with 
the recoil of the daughter nucleus. 
The $\beta$-$\alpha$ angular correlation term $p_\mp(E)$ is given 
as $-\frac{2}{3}p_\mp(E)=K(E,1)$, which is also defined by Eq.~(\ref{Eq:B2B0}).
The difference in the correlation terms between the mirror pair, 
$\delta^-_{\rm align.}=(B_2/B_0)_{\rm ^8Li}-(B_2/B_0)_{\rm ^8B}$ and
$\delta^-_{\beta\alpha}=(-2/3)(p_--p_+)$,
consists of only three terms, $b/Ac$, $g_{\rm II}/g_{\rm A}$, and $f/Ac$. 
The $b/Ac$ term is determined under the assumption that $g_{\rm II}=0$; 
$f/Ac$ is completely separated from the others 
as follows:
\begin{eqnarray}
\frac{\delta^-_{\rm align.}+\delta^-_{\beta\alpha}}{2}
&=&-\frac{2E}{3M_n}
\frac{b}{Ac} 
\label{Eq:diffsum}\\
\frac{\delta^-_{\rm align.}-\delta^-_{\beta\alpha}}{2}
&=&-\frac{2E}{3M_n}\frac{f}{Ac}\frac{E_0+2E}{\sqrt{14}E_0}.
\label{Eq:diffdiff}
\end{eqnarray}
The $c$, $b$, and $f$ terms described by reduced matrix elements as follows:
$c=g_{\rm A}\left <f||\tau^\pm \boldsymbol{\sigma} ||i\right >$, 
$b=A\left (g_{\rm M}\left <f||\tau^\pm\boldsymbol{\sigma}||i \right >+
g_{\rm V}\left <f||\tau^\pm  \boldsymbol{L}||i\right >\right )$, and
$f=2(2\pi /15)^{1/2}A M_n E_0 g_{\rm V}\left <f ||\tau^\pm r^2 Y_2(\hat{r}) ||i \right >$,
where $g_{\rm M}$ is the weak magnetism coupling constant in the weak vector current.  
$c$ is determined from the $\beta$-decay half-lives of $^8$Li and $^8$B.

The matrix elements depend on the final state energy which is broadly distributed. 
The $E_x$ dependence of $c$ and $b$ is taken into account by using $R$-matrix theory as described in Secs.~\ref{CVCprediction} and \ref{extraction}. 
The $E_x$ dependence of the others is considered as a systematic uncertainty as described in Sec.~\ref{extraction}. 

The requirement by strong CVC is that $b$ and $f$ contribute also to the electromagnetic transition from the isobaric analog state in $^8$Be. 
$b$ and $f$ are related to the isovector components of 
the M1 and E2 transition strengths, $\Gamma^{T=1}_{\rm M1}$ and $\Gamma^{T=1}_{\rm E2}$,
i.e. $b=A M_n \{6 \Gamma^{T=1}_{\rm M1}/(\alpha E_\gamma^3)\}^{1/2}$ and
$f/b=\sqrt{10/3} \delta_1$ \cite{DeBraeckeleer95}.   
Here $E_\gamma$ is the $\gamma$-ray energy, 
the fine structure constant $\alpha=1/137$,
and the M1/E2 ratio $\delta_1=(\Gamma^{T=1}_{\rm E2}/\Gamma^{T=1}_{\rm M1})^{1/2}$. 

The initial state of the analog-$\gamma$ decay splits into two isospin mixing states with $T=0$ and 1. 
In addition, the electromagnetic transitions from these states include the isoscalar and isovector components. 
Two strengths $\Gamma^{T=1}_{\rm M1}$ and $\Gamma^{T=1}_{\rm E2}$ are 
the isovector component from the state with $T=1$. 
The measurement of the $\gamma$ decay from these states and the extraction of $\Gamma^{T=1}_{\rm M1}$ and $\Gamma^{T=1}_{\rm E2}$ were performed in the previous work by De Braeckeleer {\it et al.}~\cite{DeBraeckeleer95}.

\section{Experimental}
\begin{figure}
\includegraphics{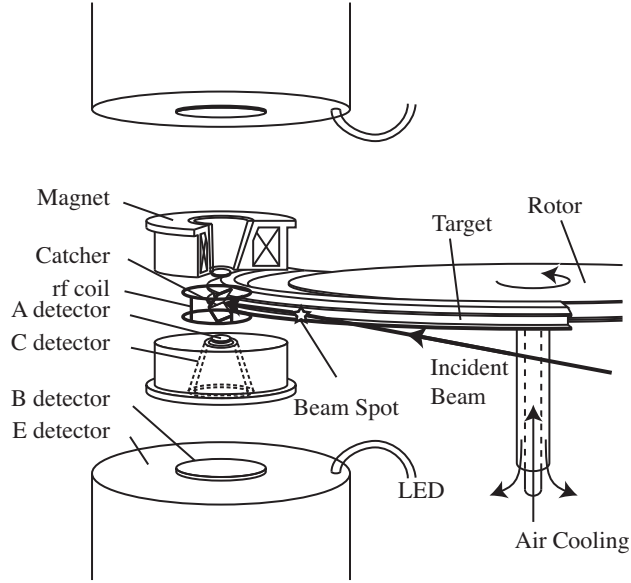}
\caption{\label{ExpSetup}
Schematic view of the experimental setup.
The rotational target with an air cooling system was used to reduce the background from the target. 
The catcher and the rf coil were placed at the center of two telescopes. 
Each plastic scintillation detector telescope consisted of  two thin $\Delta$E detectors (A and B), one veto detector ($\overline{\rm C}$), and one energy detector (E). 
}
\end{figure}

In this section, the experimental details for the alignment correlation term 
measurement is described. 
Figure \ref{ExpSetup} shows the experimental setup, 
which is essentially similar to 
the previous experiment for the alignment correlation terms 
of $^{12}$B and $^{12}$N \cite{Minamisono02-2}. 

\subsection{Production of unstable $^8$Li and $^8$B}
\label{sec:ProdPol}

The $^8$Li [$^8$B] nuclei were produced through the nuclear reactions
$^7$Li($d$,$p$)$^8$Li [$^6$Li($^3$He,$n$)$^8$B].
Hereafter, information with the parentheses represent the conditions for $^8$B.  
A Li$_2$O [enriched metal $^6$Li] target was bombarded by
a deuteron [$^3$He] beam at 3.5 MeV [4.7 MeV] 
with a typical intensity of 9 $\mu$A [40 $\mu$A].
A rotating target, which occupied one third 
of the circumference of the target rotor,
was cooled from inside the holder by a compressed air jet
in order to withstand the high-intensity $^3$He beam, which operated
at  4.7 MeV up to 40 $\mu$A. 
The pulsed beam was synchronized to the rotational period of 2.4 s.
The beam-on and beam-off times were 0.8 s and 1.6 s, respectively. 
The target material was vacuum evaporated on a backing ribbon 
made of molybdenum [phosphor bronze].
Phosphor bronze was used to reduce a Rutherford scattering of $^3$He,
which could otherwise have bombarded the recoil catcher and 
have been an origin of disturbing background activities. 
A new ion-source bottle, made of glass, was used for the $^3$He beam 
to prevent a very weak HD$^+$ molecular ion beam from mixing with the $^3$He beam. 
The HD$^+$ ion beam was formed by H$_2$ and D$_2$ gases oozing out
from the inner wall of the ion-source bottle, 
which were in turn used for the production of $p^+$ or $d^+$ beams. 

\subsection{Recoil implantation of polarized nuclei}
\label{sec:Imp}

The recoil angle of the nuclear-reaction products was selected in the range 14$^\circ$-40$^\circ$ [7$^\circ$-18$^\circ$] 
to optimize the obtained polarization.
The polarized $^8$Li [$^8$B] nuclei were implanted 
in Zn [TiO$_2$ (rutile structure)] single crystals 
by using a recoil energy of 1.7 MeV [2.3 MeV] 
obtained by the nuclear reaction.
The crystals were placed in a static magnetic field $B_0$ 
to maintain the polarization and to manipulate the spin orientation
using the $\beta$-NMR technique. 
The c axis of the single crystals was set parallel to $B_0$, 
which was 60 mT [230 mT]. 
An asymmetry of $\beta$-rays emitted from polarized nuclei 
was detected by two sets of the counter telescope 
placed in the opposite direction. 
The obtained polarization was determined  to be  $ 7.2\%$ [$5.4\%$] 
from the $\beta$-ray asymmetry by using the $\beta$-NMR technique.

\begin{figure}
\includegraphics{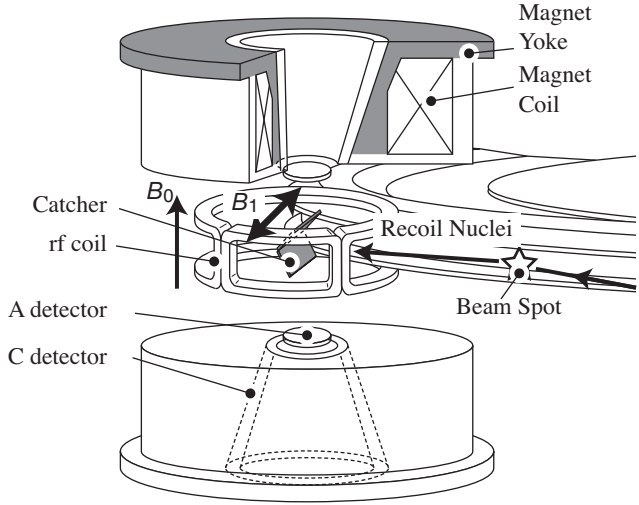}
\caption{\label{CoilChamber}
Focused view of NMR equipment.
An rf oscillating magnetic field $B_1$, which was applied by the rf coil,  
is perpendicular to the external magnetic field $B_0$. 
}
\end{figure}

The recoil catcher consisted of a pair of crystals, which were tilted 45$^\circ$ with respect to the magnetic field but in opposite directions in order to form a dog-leg shape (similar to a half-opened book) as seen from the side, as shown in Fig.~\ref{CoilChamber}. 
The implantation depth was uniformly distributed at 2.4 $\mu$m [3.1 $\mu$m] 
from the surface. 
The recoil nuclei were implanted from the inner side of the two crystals,   
making the path length and the energy loss of the $\beta$ rays in the catcher 
less sensitive to the $\beta$-emitter position. 
The thickness of crystals was $360 \pm 20$ $\mu$m for the guph Zn crystal and 
$250 \pm 20$ $\mu$m for the gdownh crystal, 
and $100 \pm 10$ $\mu$m for both TiO$_2$ crystals. 
The systematic uncertainty due to the ambiguity in the thickness 
was considered as discussed in Sec.~\ref{Syst}.

\subsection{Spin manipulation}

In order to convert the initial polarization into positive and negative
alignments with, ideally, zero polarization, 
the nuclear spin was manipulated using the NMR technique. 
The Larmor frequency for spin $J=2$ nucleus splits into 
four resonance frequencies 
because of  hyperfine interaction 
between the electric quadrupole moment $Q$ of the implanted nucleus and
the electric field gradient (EFG) at an implantation site in the crystal.
EFG is defined by $V_{ii}={d^2V}/{di^2}$, 
where $i$ is the principal axes of EFG, i.e., $X,Y$, and $Z$, $V_{XX}+V_{YY}+V_{ZZ}=0$,
and $|V_{XX}|\le|V_{YY}|\le|V_{ZZ}|$. 
Therefore, once principal axes are chosen, EFG is given by two parameters $q=V_{ZZ}$ and 
$\eta=(V_{XX}-V_{YY})/V_{ZZ}$. 
The resonance frequency between two neighboring magnetic substates, 
$(m-1)\leftrightarrow m$, is given in \cite{Abragam} as
\begin{equation}
\nu_{m-1\leftrightarrow m}=\nu_L -
\frac{\nu_Q}{4}(3\cos^2\theta -1 + \eta \sin^2\theta \cos 2\phi)(2m-1),
\end{equation}
where $\nu_L$ is the Larmor frequency, $\nu_Q=eqQ/4h$, 
and $\theta$ and $\phi$ are the Euler angle between 
the principal axes of EFG and the external magnetic field, respectively.

\begin{figure}
\includegraphics{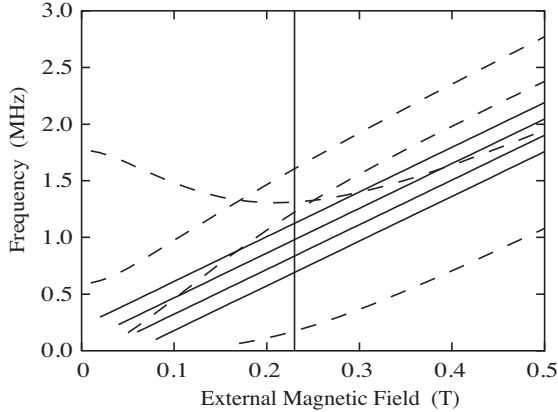}
\caption{\label{fvsH0}
External magnetic field dependence of the resonance frequencies 
of $^8$B in TiO$_2$. 
The solid and dashed lines denote the frequencies of major and minor sites, respectively. 
}
\end{figure}

Populations of two neighboring magnetic substates can be manipulated
independently by applying an rf oscillating magnetic field 
at each frequency.
EFG at implantation sites in crystals has been studied by the $\beta$-NMR technique \cite{eqQ1,eqQ2,eqQ3}.
The number of possible implantation sites is one for $^8$Li in Zn and 
two for $^8$B in TiO$_2$.
The relative populations are 90\% for $^8$B implanted 
in the major site of TiO$_2$ and 10\% for that in the minor site.
$\nu_Q$ and $\eta$ have been determined as $\nu_Q=+8.4\pm0.5$ kHz and $\eta=0$ 
for the implantation site of $^8$Li in Zn \cite{eqQ1}
and as $\nu_Q=+144.5\pm0.6$ kHz and $\eta<0.03$ 
for the major implantation site of $^8$B in TiO$_2$ \cite{eqQ3}.
Because of a small population for the minor site, 
it was difficult to detect a $\beta$-NQR signal for $^8$B in the minor site. 
$\nu_Q$ and $\eta$ at the minor site of $^8$B in TiO$_2$ was evaluated 
as $\nu_Q=+1185\pm8$ kHz and $\eta=0.020\pm0.006$
from $\nu_Q$ at the minor site of $^{12}$B \cite{eqQ2}
and the ratio of the $Q$ moments of $^8$B and $^{12}$B \cite{eqQ3}. 
The directions of $q$ at an implantation site of  $^8$Li in Zn and 
at the major site of $^8$B in TiO$_2$ \cite{eqQ1,eqQ3}
were parallel to the c axis of the crystals, i.e., $\theta=0$, 
thus giving four frequencies split at regular intervals. 
For the minor site of $^8$B in TiO$_2$,
the direction of $q$ was inclined at 106$^\circ$ 
relative to the $\langle100\rangle$ axis on the $(001)$ plane and 
the direction of $V_{\rm YY}$ was parallel to the c axis of the crystal \cite{eqQ2},
i.e., $\theta=\phi=90^\circ$, as the c axis was parallel to $B_0$.
Frequencies for $^8$B at the major and the minor sites 
are shown as a function of $B_0$ in Fig.~\ref{fvsH0}.
Frequencies for the major site were isolated from those of the minor site
only for the experimental condition of $B_0=230$ mT.
Under this condition, only the nuclear spin of $^8$B implanted in the major site 
can be manipulated. 
The $\beta$-ray angular distribution from the unmanipulated $^8$B 
in the minor site was stable. 
Because the alignment correlation term was derived
from the dependence of the $\beta$-ray angular distribution 
on the degree of the alignment,
the effect of $^8$B in the minor site was canceled. 

\begin{figure}[ht]
\includegraphics{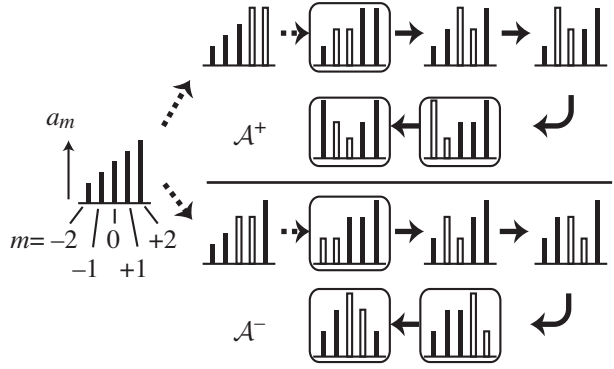}
\caption{\label{AlignProd}
Spin-aligning procedure for $^8$Li. 
The change in the populations, $a_m$, of the magnetic substate are shown. 
The spin manipulations with the AFP and depolarization methods of the NMR technique are
denoted by the solid and dashed arrows, respectively.
The two open bars in each orientation show the manipulated populations.
The upper and lower parts show the production procedure for  
the positive and negative alignments, respectively. 
The polarizations of the three orientation patterns framed 
by the separate squares were measured to determine the alignment. 
The timing program for the measurement is shown in Fig.~\ref{AlignProd2}. 
}
\end{figure}

\begin{figure}[ht]
\includegraphics[scale=0.8]{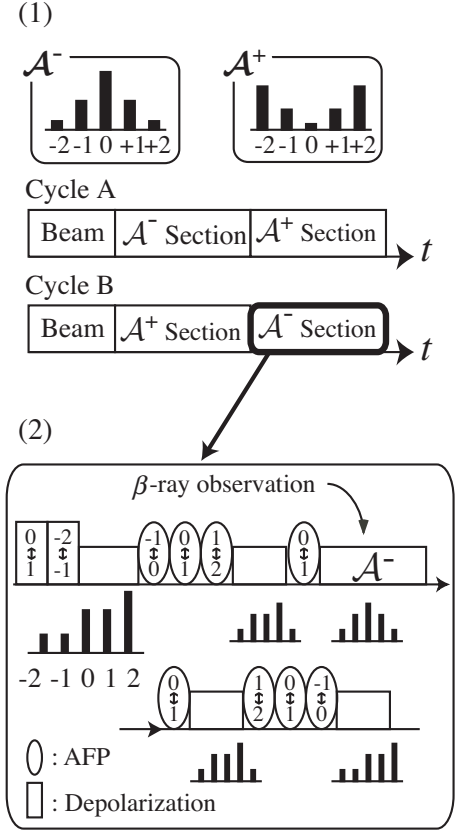}
\caption{\label{AlignProd2}
Timing program for the spin-aligning process. 
In each cycle, the positive and negative alignments were produced as shown in (1). 
The timing program for the spin manipulation and the $\beta$-ray-angular-distribution measurement
are shown in (2). 
The ellipses and squares with two numbers show the spin manipulation with
the AFP and depolarization methods, respectively,  
used for the relevant transition between the magnetic substate
nominally shown by the two numbers. 
The $\beta$-ray angular distribution was observed at the half-height squares. 
The alignment was converted back to the polarized form to check the spin manipulation 
and to measure alignment relaxation. 
}
\end{figure}

The spin-aligning procedure for spin $J=2$ was newly developed as part of our study. 
Figure \ref{AlignProd} shows
the schematic aligning procedure using $^8$Li as an example. 
Immediately after the pulsed beam was stopped, 
the nuclear spin was manipulated by applying two kinds of 
$\beta$-NMR technique, 
the adiabatic fast passage (AFP) and the depolarization methods. 
The populations between the two neighboring magnetic substates 
were interchanged by the AFP method and equalized by the depolarization method.
To convert a positive polarization to a positive alignment ${\cal A}^+$,
the populations in $m=+2$ and $+1$, as well as in $m=-1$ and $0$ 
were first of all equalized using the depolarization method. 
Following this, the positive alignment was produced by sequentially applying 
the AFP method four times, by which the populations  
between $m=+1$ and $0$, $m=-1$ and $0$, 
$m=-2$ and $-1$, as well as $m=-1$ and $0$ were interchanged. 
A negative alignment was produced immediately 
after the beam was stopped in the next beam-count cycle
following a similar procedure 
applied to the magnetic substates as shown in the ${\cal A}^-$ part 
of Fig.~\ref{AlignProd}.
For $^8$B, an opposite sign of alignment was produced 
using the same procedure as for $^8$Li, 
because the polarization initially obtained for $^8$B was negative
while the other parameters, i.e., the direction of the holding magnetic field and
the field gradient, were similar. 
The alignment was converted back to a polarization 
to check the spin manipulation 
and to measure the relaxation time of the alignment. 
Subsequently, in the same beam-count cycle, 
the polarization was converted to an alignment with the opposite sign, 
as shown in Fig.~\ref{AlignProd2}. 
This method of data acquisition using the present timing program 
removed the systematic uncertainty 
due to beam fluctuation as described in Sec.~\ref{analysis:ACT}. 

\subsection{Beta-ray energy spectra}

The $\beta$ rays were detected 
by two sets of plastic scintillation counter telescopes 
placed above ($\theta_{I\beta}=0^\circ$) and below ($180^\circ$) the crystal
as shown in Fig.~\ref{ExpSetup}. 
Each telescope consisted of 
two thin $\Delta$E (A and B) detectors  
of 12 mm$\phi\times$0.5 mm and 55 mm$\phi\times$1 mm,
one $\beta$-ray energy (E) detector of 160 mm$\phi\times$120 mm, and 
one cone-shaped veto (${\overline {\rm C}}$) detector.
The ${\overline {\rm C}}$ detector was used to reject the $\beta$ rays scattered at the magnet.
A typical counting rate of $\beta$ rays from $\beta$ emitters 
stopped in the catcher was 4 kcps (1.5 kcps).

\begin{figure}
\includegraphics{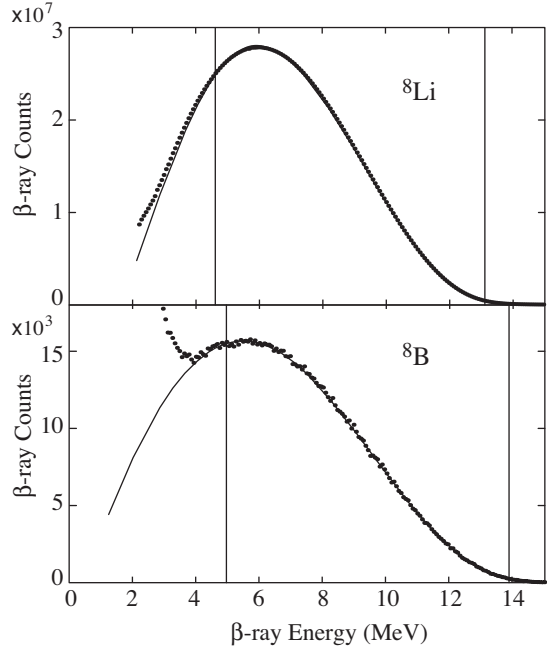}
\caption{\label{BetaSpectra}
Typical $\beta$-ray energy spectra for $^8$Li (upper) and $^8$B (lower). 
The dots are the experimental data and the solid curves are the 
best-fit lines. 
The energy region lying between the two vertical lines in each energy spectrum 
shows the region used for the line fitting.  
The background $\beta$ rays in the low energy region of the $^8$B spectrum were from $^{15}$O. 
}
\end{figure}

The energy spectra of $\beta$-rays emitted 
from purely aligned $^8$Li and $^8$B are 
shown in Fig.~\ref{BetaSpectra}.
The gain in the analog signal
was stabilized using the standard light pulse 
from a light-emitting-diode (LED) pulser 
whose the circuit was maintained at a constant temperature.

\begin{figure}
\includegraphics{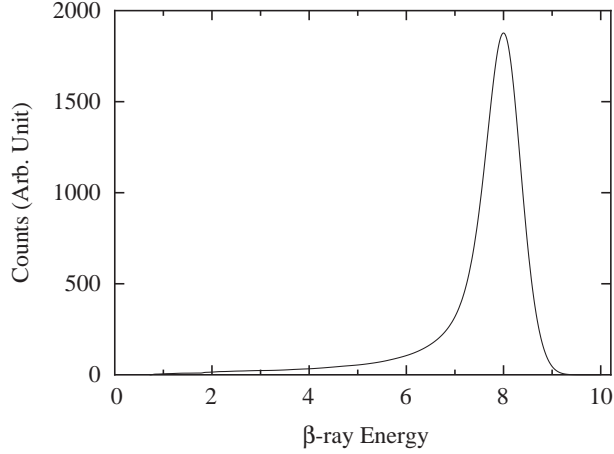}
\caption{\label{Response}
Response function of the E detector for $\beta^-$ ray  with 8 MeV. 
The horizontal axis has been rescaled from the simulated energy deposit in the E detector
to the $\beta$-ray energy just after the emission from the nucleus. 
}
\end{figure}

The energy deposit in the E detector for a monoenergetic $\beta$ ray was obtained 
by a Monte Carlo simulation with the EGS4 code \cite{EGS}. 
The detector telescopes, the catcher of the reaction products $^8$Li and $^8$B, the catcher holder, 
and the vacuum chamber near the $\beta$-ray window were 
arranged in the simulation. 
The distribution of the reaction products on the catcher was given using the reaction kinematics. 
The response function was obtained by convoluting 
the deposit function with a detector resolution as shown in Fig.~\ref{Response}. 
The resolution of the Gaussian function was determined 
by the $\chi^2$ fitting of the $\beta$-ray energy spectra of $^8$Li and $^8$B 
with $\sigma=\sigma_0\sqrt{E_{dep}}$, where $\sigma_0=0.10\pm0.02$ (MeV)$^{1/2}$. 
Here, $E_{dep}$ is the energy deposit in the E detector, which was observed, 
whereas the alignment correlation term needed to be extracted 
as a function of the $\beta$-ray energy 
just as it was emitted from the nucleus. 
The peak position of the energy deposit for monoenergetic $\beta$ rays was scaled to the incident energy of the $\beta$-ray. 
The $\beta$-ray energy spectrum for the $\chi^2$ fitting was obtained 
by convoluting the $\beta$-ray continuous energy spectrum 
with the response function of the monoenergetic $\beta$ ray. 

The $\beta$-ray energy was scaled by determining 
the end-point energies of several $\beta$-emitters,
which were $^8$Li itself, $^{28}$Al($E_0=2.86$ MeV), $^{20}$F(5.39 MeV), 
and $^{12}$B (13.37 MeV) for the $^8$Li experiment, 
and $^8$B itself, $^{15}$O(1.73 MeV), $^{20}$F(5.39 MeV), and 
$^{12}$N(16.32 MeV) for the $^8$B experiment.

\section{Analysis}

First, the determination of the degree of polarization and alignment, and then that of the alignment correlation terms are described.  The evaluation of corrections and systematic uncertainties follows. 

\subsection{Degree of polarization and alignment}
\label{degPolAlign}
The polarization was determined from the $\beta$-ray asymmetry 
where $\beta$ rays from 5 to 13 MeV were used. 
The $\beta$-ray angular distribution from the polarized nuclei is given by
$W(\theta_{I\beta}) \propto B_0(E)+ B_1(E) {\cal P}\cos\theta_{I\beta} \propto 1 + A_s {\cal P} \cos\theta_{I\beta}$, where ${\cal P}$ is the degree of the polarization.
The asymmetry parameter, $A_s=B_1(E)/B_0(E)$, 
has an energy independent-main term and an energy-dependent higher order term.  
For the determination of the degree of the polarization and alignment, 
$A_s$ was approximated as $- 1/3$ for $^8$Li and $+1/3$ for $^8$B. 
The effect on the alignment correlation term by the higher order term of $A_s$ 
was corrected as described in Sec.~\ref{CorrHigher}. 
The counting ratio of the top and bottom telescopes, 
i.e. $\theta_{I\beta}=0^{\circ}$ and $180^\circ$, 
was caused by the $\beta$-ray asymmetry from the polarization ${\cal P}$ as well as 
the possible geometrical asymmetry $g$ resulting from the geometrical 
misalignment between two telescopes. 
This ratio is expressed as 
$R_\beta=W(0^\circ)/W(180^\circ)=g (1+A_s{\cal P})/(1-A_s{\cal P})$.
To determine $g$, 
the polarization was inverted by applying a series of 10 AFPs.  
The measured counting ratios for the initial polarization $R_{{\cal P}+}$, 
the inverted polarization $R_{{\cal P}-}$, and the twice inverted polarization
$R_{{\cal P}++}$ are given by
\begin{eqnarray}
R_{{\cal P}+}&=&g (1+A_s{\cal P}_0)/(1-A_s{\cal P}_0)\\
R_{{\cal P}-}&=&g (1+\alpha A_s{\cal P}_0)/(1-\alpha A_s{\cal P}_0)\\
R_{{\cal P}++}&=&g (1+\alpha^2 A_s{\cal P}_0)/(1-\alpha^2A_s{\cal P}_0).
\end{eqnarray}
From these equations, 
the initial polarization, ${\cal P}_0$, $g$ 
and the polarization inversion efficiency $\alpha$ were deduced, 
as shown in Table \ref{ResultSpinMani}.
The inversion efficiency $\eta$ for the populations 
between the two magnetic substates by one AFP 
were determined from the relationship between $\alpha$ and $\eta$, 
$\alpha\approx 4-5\eta$.
The relaxation time of the polarization $T_1$ was determined 
from the time spectrum of polarization. 
These parameters are given in Table~\ref{ResultSpinMani}.

\begin{figure}
\includegraphics{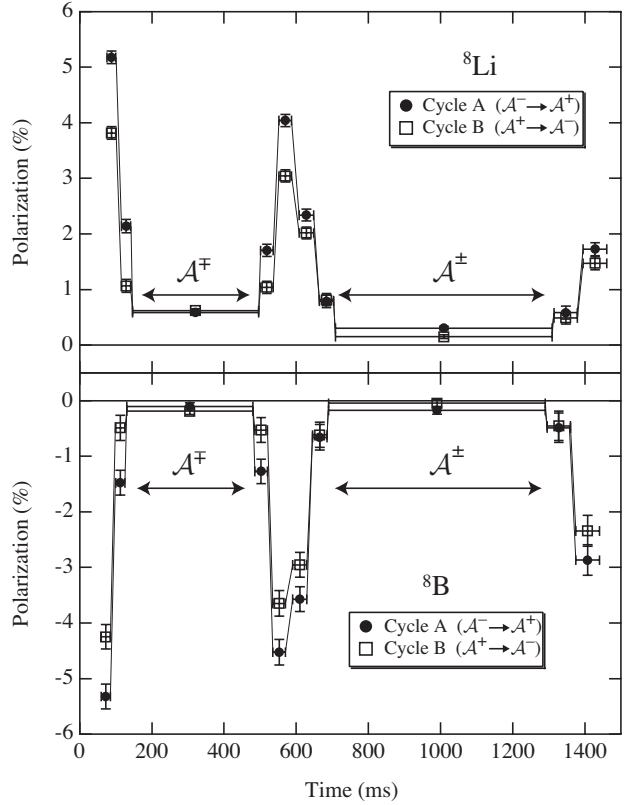}
\caption{\label{PolChaA}
Polarization change in the timing program for the spin-aligning process. 
The filled circles and the open squares are for the cycles A and B 
in Fig.~\ref{AlignProd2}, respectively. 
The beam was chopped and was stopped for the cycle at the time 0.}
\end{figure}

The extraction of the degree of alignment from the negative alignment section shown in 
Fig.~\ref{AlignProd} was performed as follows. 
The $\beta$-ray asymmetry of the three orientations
shown in Fig.~\ref{AlignProd} was observed
during the aligning process. 
The polarization was determined from the measured asymmetry and $g$, 
as shown in Fig.~\ref{PolChaA}.  
The population of the magnetic substate at the first orientation is given by 
 $[a_{-2},a_{-1},a_0,a_{+1},a_{+2}]=[r (1-\epsilon_1),r (1+\epsilon_1),s(1-\epsilon_2),s(1+\epsilon_2),t]$. 
The parameters  $r, s$, and $t$ satisfy the relation $2r+2s+t=1$.  
$\epsilon_1$   and $\epsilon_2$  are the parameters describing incompleteness in the depolarization method for the two different frequencies. 
These two parameters  yielded a small residual polarization at the pure alignment section. 
 The polarization of the first orientation is given by 
 ${\cal P}_{\rm 1st} =  \frac{1}{2} \{ r (\epsilon_1-3) + s (\epsilon_2 +1 ) + 2 t \}$. 
The population after the spin manipulation using the AFP method, for example, between $m=+2$ and $m=+1$ is given as a matrix: 
\begin{equation}
\begin{pmatrix}
1-\eta & \eta & 0 & 0 & 0 \\
\eta & 1-\eta & 0 & 0 & 0 \\
0 & 0 & 0 & 0 & 0 \\
0 & 0 & 0 & 0 & 0 \\
0 & 0 & 0 & 0 & 0 
\end{pmatrix}
\begin{pmatrix}
a_{+2}\\
a_{+1}\\
a_{0}\\
a_{-1}\\
a_{-2}
\end{pmatrix},
\end{equation}
where $\eta \approx 1$. 
The spin manipulation shown in Fig.~\ref{AlignProd}-(2) can be described as the product of the matrices.  
Therefore, the population at each orientation is given by $r, s, t, \eta, \epsilon_1$, and $\epsilon_2$. 
The polarization at the second orientation is given by
${\cal P}_{\rm 2nd} \approx \frac{1}{2} \{ 4 r \epsilon_1 + s (\epsilon_2 - 1 ) +  t 
+ (1-\eta )(- 6 r (\epsilon_1 + 1 ) - s (\epsilon_2 - 5) + t )  \}$ 
under the approximation, up to the first order, that $(1-\eta) \ll 1$.  
The pure alignment is produced at the third orientation. 
The residual polarization is given by 
${\cal P}_{\rm 3rd} \approx  \frac{1}{2} \{ 4 r \epsilon_1 + 2 s \epsilon_2 
+ (1-\eta )( - 6 r (\epsilon_1 + 1 ) - 3 s (\epsilon_2 - 1) + 3 t )  \}$. 
$\eta$ was determined from the measurement of $R_{\cal P^+}, R_{\cal P^-},$ and $R_{\cal P^{++}}$.  
Therefore, the number of free parameters is three
by assuming $\epsilon_1= \epsilon_2$ and giving the relation $2r+2s+t = 1$.  
All the population parameters were determined  
from the polarization change of the three orientations. 

The alignment in the third orientation can be calculated from the population parameters using
${\cal A} \approx \frac{1}{2}\{ -2 + 8r + 2s +(1-\eta) (-10 r (\epsilon_1 + 1) - s (\epsilon_2 -5) + 5t
\}$.
This equation gives the alignment prior to the alignment section. 
In order to consider the alignment relaxation in the crystal, 
the alignments prior to and after the alignment section were determined  
from the polarization change before and after the alignment section, respectively.
Then the effective alignment and the relaxation time of the alignment
were deduced.
Using a different assumption that $\epsilon_1= 10 \epsilon_2$ or $\epsilon_1= 0.1 \epsilon_2$, 
the systematic uncertainty  was estimated. 
The change in the alignment was less than the statistical uncertainty. 
The results for the spin manipulation are summarized in Table \ref{ResultSpinMani}.

\begin{table}[hbt]
\caption{\label{ResultSpinMani}
Results of the spin manipulation. 
${\cal P}_0$ is the initial polarization.  
${\cal A}^{\mp}_{\rm 1/ 2}$ is the alignment, where subscripts 1 and 2 indicate the first and second halves of the timing program, respectively, 
and  the superscript is the sign of the alignment. 
$\Delta {\cal A}_{1+2}$ is the sum of the absolute value  of the alignments. 
$T_1$ and $T_{\cal A}$ are the relaxation times of the polarization and the alignment, respectively. 
$\alpha$ and $\eta$ are the efficiencies of the polarization inversion and 
the population inversion between the two neighboring magnetic substates. 
$\epsilon$ is a parameter of the incompleteness of the depolarization. 
}
\begin{ruledtabular}
\begin{tabular}{ccc}
&$^8$Li&$^8$B\\
\hline
${\cal P}_0$ (\%)&$7.18\pm0.10$&$5.42\pm0.19$\\
${\cal A}_1^+$ (\%)&$+3.96\pm0.20$&$+4.9\pm0.4$\\
${\cal A}_1^-$ (\%)&$-4.93\pm0.20$&$-5.6\pm0.4$\\
${\cal A}_2^+$ (\%)&$+2.29\pm0.19$&$+3.9\pm0.4$\\
${\cal A}_2^-$ (\%)&$-1.91\pm0.19$&$-3.2\pm0.4$\\
$\Delta {\cal A}_{1+2}$ (\%)&$13.1\pm0.4$&$17.7\pm0.8$\\
$T_1$ (s)&$13.0\pm1.6$&$13\pm4$\\
$T_{\cal A}$ (s)&$2.0\pm0.7$&$2.2\pm1.2$\\
$\alpha$ (\%)&$-85.5\pm0.3$&$-94.8\pm0.9$\\
$\eta$ (\%) & $97.09\pm0.07$&$98.95\pm0.18$\\
$\epsilon$ $(10^{-3})$ &$4.4\pm0.2$&$-0.9\pm0.4$\\
\end{tabular}
\end{ruledtabular}
\end{table}

\subsection{Alignment correlation term}
\label{analysis:ACT}
The alignment correlation term was obtained from the ratio of counts, 
$R(E)= N(E,d{\cal P}^+,{\cal A}^+)/N(E,d{\cal P}^-,{\cal A}^-)$,
at the positive and negative alignment sections. 
${\cal A}$ and $d{\cal P}$ are the alignment and the residual polarization 
at the alignment section, respectively. 
The signs given by the superscript in ${\cal A^\pm}$ and $d{\cal P^\pm}$ are 
the alignment signs.
The counts are proportional to the $\beta$-ray angular distribution 
as expressed by $N(E,d{\cal P},{\cal A})\propto B_0(E)[1\pm (B_1(E)/B_0(E)) d{\cal P}+
(B_2(E)/B_0(E)) {\cal A})]$,
where the upper and lower signs are for the top and bottom telescopes, respectively.

For the first half of the cycles A and B shown in Fig.~\ref{AlignProd2}, 
the counting ratio $R_1(E)$ is given by
\begin{equation}
R_1(E)
=\frac{T^+ N(E,d{\cal P}^+_1,{\cal A}^+_1)}
{T^- N(E,d{\cal P}^-_1,{\cal A}^-_1)},
\end{equation}
where the values with subscript 1, such as ${\cal A}^+_1$, 
are for the first half. 
$T^-$ and $T^+$ are the beam-current integral for the cycles A and B, respectively.
The alignment correlation term was derived using the well-approximated formula as
\begin{equation}
\label{Ratio}
R_1(E)\approx \frac{T^+}{T^-}\left\{1\pm \frac{B_1(E)}{B_0(E)}d{\cal P}_1+
\frac{B_2(E)}{B_0(E)}\Delta {\cal A}_1\right\},
\end{equation}
where the upper and lower signs are for the top and bottom telescopes, respectively. 
$d{\cal P} _1= d{\cal P}_1^+ -d{\cal P}_1^-$ and
$\Delta {\cal A}_1 = {\cal A}_1^+ -{\cal A}_1^-$.
The ratio of $T^+$ and $T^-$ caused a spurious $\beta$-ray asymmetry in $R_1(E)$.
The counting ratio $R_2(E)$ at the second half of the cycles A and B is given by 
\begin{eqnarray}
R_2(E)&=&
\frac{T^- N(E,d{\cal P}^+_2,{\cal A}^+_2)}
{T^+ N(E,d{\cal P}^-_2,{\cal A}^-_2)} \nonumber \\
&=&  \frac{T^-}{T^+}\left\{1\pm \frac{B_1(E)}{B_0(E)}d{\cal P}_2+
\frac{B_2(E)}{B_0(E)}\Delta {\cal A}_2\right\},
\end{eqnarray}
where the values with subscript 2, such as ${\cal A}^+_2$, 
are for the second half. 
In the double ratio $R_1(E)R_2(E)$, $T^+$ and $T^-$ are canceled 
as 
\begin{eqnarray}
R_1(E)R_2(E)&=&
\frac{N(E,d{\cal P}^+_1,{\cal A}^+_1)}{N(E,d{\cal P}^-_2,{\cal A}^-_2)}
\frac{N(E,d{\cal P}^+_2,{\cal A}^+_2)}{N(E,d{\cal P}^-_2,{\cal A}^-_2)} \nonumber \\
&\approx&1\pm\frac{B_1(E)}{B_0(E)}d{\cal P}_{1+2} \nonumber \\
&&+\frac{B_2(E)}{B_0(E)}\Delta{\cal A}_{1+2}, \label{R1R2}
\end{eqnarray}
where $d{\cal P}_{1+2}=d{\cal P}_1+d{\cal P}_2$ and 
$\Delta{\cal A}_{1+2}=\Delta{\cal A}_1+\Delta{\cal A}_2$. 
The alignment correlation terms were extracted 
from the simple average of the double ratios $R_1(E)R_2(E)$ for the top and bottom telescopes
so that the influence of the residual polarization was canceled.

\subsection{Corrections}
\label{Corrections}
In the extraction procedure for the alignment correlation terms described above, 
the $\beta$-ray angular distribution for $^8$Li is given by 
\begin{equation}
W(E)\propto pE(E_0 -E ) \left \{ 1 \mp \frac{1}{3} {\cal P} +  \frac{B_2(E)}{B_0(E)} {\cal A}\right \},
\end{equation} 
where the upper and lower signs are for telescopes with $\theta_{I \beta}= 0^{\circ}$ and $180^{\circ}$, respectively, 
instead of the following: 
\begin{eqnarray} 
W(E, \theta_{I \beta})&\propto& pE(E_0 -E )   \nonumber  \\
&&\left \{ 1 + \frac{B_1(E)}{B_0(E)} \frac{p}{E} {\cal P} P_1(\cos(\theta_{I \beta})) \right . \nonumber \\
&&\left. +  \frac{B_2(E)}{B_0(E)} \left (\frac{p}{E}\right )^2 {\cal A} P_2(\cos(\theta_{I \beta}))\right \}. 
\end{eqnarray} 
The correction for the $P_1(\cos(\theta_{I \beta}))$, and $P_2(\cos(\theta_{I \beta}))$ is 
given in Sec.~\ref{solid}. 
The corrections for $(p/E)$, $(p/E)^2$, and $B_1(E)/B_0(E)$ are given in Sec.~\ref{CorrHigher}.

\subsubsection{Solid angle of $\beta$-ray telescope}
\label{solid}
The polarization and alignment correlation terms in the $\beta$-ray angular distribution 
are proportional to the Legendre polynomials $P_1(\cos(\theta_{I \beta}))$ and $P_2(\cos(\theta_{I \beta}))$, i.e., the
$\cos\theta_{I \beta}$ and $\frac{3}{2}(\cos^2\theta_{I \beta}-1/3)$ terms, respectively. 
$R_1(E) R_2(E)$ in Eq.~(\ref{R1R2}) includes $B_2(E)/B_0(E)$, so 
the $\frac{3}{2}(\cos^2\theta_{I \beta}-1/3)$ contribution should be corrected. 
$\Delta{\cal A}_{1+2}$ was determined from the degree of polarization,  
so the $\cos\theta_{I \beta}$ contribution should be corrected.  
In order to take the finite solid angle of the detector into account, 
the detection efficiency as a function of  $E$ and $\theta_{I \beta}$ was simulated using the EGS4 code. 
The correction for the solid angle was evaluated by convoluting the simulated efficiency,
as shown in Fig.~\ref{Corr}. 

\subsubsection{Higher order term in the polarization and alignment correlation terms}
\label{CorrHigher}
The $p/E$ term and the $ B_1(E)/B_0(E)$ term in the polarization correlation term were assumed 
to be 1 and  $\mp1/3$, where the upper and lower signs are for $^8$Li and $^8$B, respectively,
when the polarization was determined from the $\beta$-ray asymmetry. 
The correction for the polarization is independent of energy, 
because the polarization was determined from the total count from 5 to 13 MeV. 
The correction for the $p/E$ term was 0.9972 and 0.9973 for $^8$Li and $^8$B, respectively. 
The $ B_1(E)/B_0(E)$ term is given in \cite{Holstein74} as
\begin{eqnarray}
\frac{B_1(E)}{B_0(E)} &=& \mp \frac{1}{3} \left [1
+ \frac{E}{3M_n}\left (\frac{1}{A}\pm\frac{b}{Ac}-\frac{d_{\rm I}}{Ac}\right )\right . \nonumber \\
&&
-\frac{\sqrt{21}}{4}\left\{ 
\pm \frac{f}{Ac}\frac{4E+E_0+4E^2/E_0}{3M_n}
\right . \nonumber \\
&&\left . \left .
-\frac{j_2}{A^2c}\frac{8E^2-5E E_0}{2M_n^2}
\right\}
\right ]. 
\label{B1B0}
\end{eqnarray}
To avoid the large systematic uncertainty from the $j_2/A^2c$ term, 
the correction factor was evaluated using the product of the correction factor at $\frac{5}{8}E_0$
and the ratio of the value at $\frac{5}{8}E_0$ to
the averaged value from 5 to 13 MeV. 
The ratio was determined, from the observed energy dependence of the polarization correlation term, 
to be $0.983\pm0.007$ for $^8$Li and $1.013\pm0.014$ for $^8$B. 
The correction factor at $\frac{5}{8}E_0$ was self-consistently evaluated using iteration
to be $0.98\pm 0.03$ for $^8$Li and $0.99\pm0.03$ for $^8$B
from the matrix elements $b/Ac$, $d_{\rm I}/Ac$, $f/Ac$, and $j_2/A^2c$, 
which in the present study were determined 
from the alignment correlation terms and the $\beta$-$\alpha$ angular correlation terms.  
The uncertainty of this correction included the uncertainty of the matrix elements and
a 100\% uncertainty of the higher order contribution from $f/Ac$, 
thus implying a severe evaluation. 
Accordingly, the correction  factor for $ B_1(E)/B_0(E)$ was $0.96 \pm 0.03$ for $^8$Li and $1.00\pm 0.03$ for $^8$B. 

The $(p/E)^2$ term in the alignment correlation term is assumed to be 1
for the first-order analysis. 
The evaluated correction factor for the $(p/E)^2$ term is shown in Fig.~\ref{Corr}. 

\subsubsection{Detector response}

The observed alignment correlation term includes the contribution from the neighboring energy region
to some extent because of the finite detector resolution and the low-energy tail component of the detector response, as shown in Fig.~\ref{Response}. 
The correction factor was evaluated self-consistently 
using the known detector response and the alignment correlation term, as shown in Fig.~\ref{Corr}.
Here, the alignment correlation term was approximated by a quadratic curve, $c_1 E+c_2 E^2$, 
with two parameters $c_1$ and $c_2$. 
The correction factor for $^8$B from 6 to 12 MeV was close to 1.0
because the alignment correlation term was almost constant and the influence 
of the different energy was small.

\subsubsection{ Background}
\begin{figure}
\begin{center}
\includegraphics{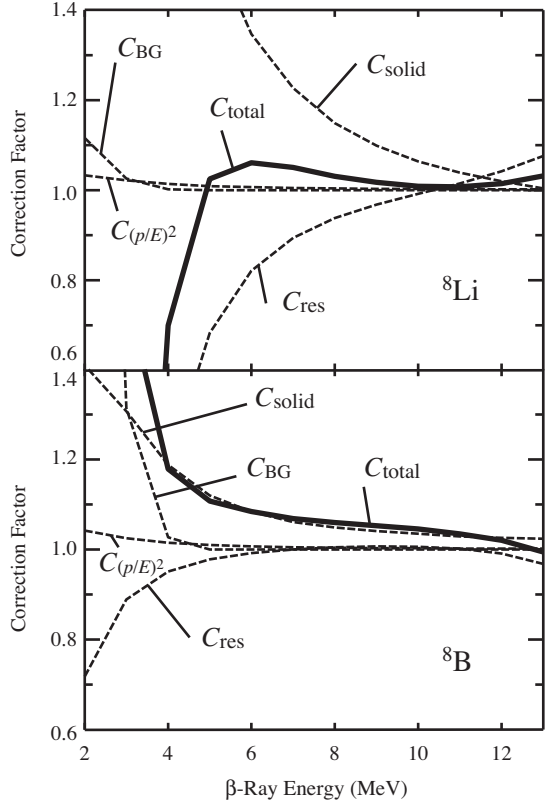}
\caption{\label{Corr}
Energy dependent correction factors and total correction for 
$^8$Li (upper) and  $^8$B (lower).
The correction factors for the solid angle, $C_{\rm solid}$, 
the detector response, $C_{\rm res}$, 
the background, $C_{\rm BG}$, and 
the $(p/E)^2$ term in the $\beta$-ray angular distribution, $C_{(p/E)^2}$,
are shown.
The total correction factor, $C_{\rm total}$ also includes the energy independent correction factors
for the $(p/E)$ and $B_1/B_0$ terms in the $\beta$-ray angular distribution. 
}
\end{center}
\end{figure}

The main backgrounds for $^8$Li and $^8$B below 4 MeV were
$^{17}$F($T_{1/2} = 64.5$ s, $Q_{EC} = 2.76$ MeV) and 
$^{15}$O($T_{1/2} = 122$ s, $Q_{EC} = 2.75$ MeV), respectively.
The correction for the background is also shown in Fig.~\ref{Corr}. 
The systematic uncertainty in the alignment correlation term 
was estimated by assuming 20\% ambiguity in the background fraction. 

\subsection{Systematic uncertainties}
\label{Syst}

In this subsection, the systematic uncertainties of the alignment correlation term are 
described. They are also summarized in Table \ref{Syst9MeV}.

\subsubsection{Polarization relaxation and rank-three spin orientation}

The polarization and alignment were relaxed as a function of time. 
While the alignment relaxation during the alignment correlation term measurement 
was taken into account in the procedure of the alignment extraction,  
the relaxation during conversion from the polarization to alignment 
was evaluated as a systematic uncertainty. 

The $\beta$-ray asymmetry was caused by the odd rank of the spin orientation.
The degree of the rank-three spin orientation was determined from 
the population parameters similar to the case of the degree of alignment, 
as described in Sec.~\ref{degPolAlign}.  
The polarization was evaluated by taking the degree of the rank-three orientation into account.
The effect on the polarization was considered as a systematic uncertainty.

\subsubsection{Uncertainty in the correction factor resulting from self-consistent evaluation}

The correction factor for the detector response
was self-consistently evaluated using the results of the alignment correlation term. 
The energy dependence of the alignment correlation term 
was estimated as a quadratic curve without a constant term. 
The statistical uncertainty of the quadratic curve was propagated to the systematic uncertainty. 

The correction factor for $B_1(E)/B_0(E)$ was evaluated 
using the matrix elements, such as  $d_{\rm I}/Ac$. 
The systematic uncertainty for this correction factor has been discussed in Sec.~\ref{Corrections}.

\subsubsection{Relative positions of the beam, recoil catcher, and telescope}

The implanted recoil nuclei distributed widely on the catcher except 
for the part in the shadow part of the collimator. 
The relative positions of the beam spot and the catcher
were able to change the distribution of the reaction products on the catcher. 
The beam spot was tuned using a fluorescent target  
with accuracy 0.5 mm and 1 mm in the horizontal and vertical directions, respectively. 
The relative position of the catcher and the $\Delta$E detector had the potential to change
the solid angle of the telescope and $\beta$-ray efficiency. 
The uncertainty of this relative position was 2 mm. 
The correction factors  were evaluated 
by using the detector response simulated for different conditions using the EGS4 code.
The change in the alignment correlation terms because of these two uncertainties was considered to be the systematic uncertainty. 

\subsubsection{Detector response function}

The reliability of the low-energy tail in the simulated response function of 
a mono-energetic $\beta$-ray was evaluated. 
The low-energy tail was mainly caused by the energy loss straggling in the material 
between the positions of the $\beta$-ray emitter and E detector. 
The largest uncertainty was due to the catcher thickness. 
The relative uncertainty of the thickness was 10\% for all the crystals. 

The reliability of the simulated low-energy tail has been studied experimentally \cite{Tanaka}. 
$^{12}$B and $^{12}$N were produced as ƒÀ emitters. 
The ƒÀ-ray energy was selected via a dipole magnet. 
The shape and amount of low-energy tail were confirmed to within 20\% statistical uncertainty.

The catcher thickness uncertainty of 10\% and the simulation reliability of 20\% for the low-energy tail 
were simulated simultaneously by varying the crystal thickness by 30\% in the EGS4 simulation. 
The correction factors were evaluated by simulating the detector response at a crystal thickness varied by 30\%. The systematic uncertainty in the alignment correlation terms was evaluated using these correction factors. 

The detector resolution was determined 
from that reproducing the most complete experimental $\beta$-ray spectra of $^8$Li and $^8$B. 
The uncertainty of the counter resolution was 20\%.  
The correction factors were evaluated using counter resolutions both the 20\% larger and 20\% smaller than the most probable resolution. 
The systematic uncertainty in the alignment correlation terms was evaluated using these correction factors. 

\subsubsection{Energy scaling, gain fluctuation, and pileup}

The systematic uncertainty due to the energy-scale uncertainty $\delta E$ was evaluated
using $\frac{d}{dE} (B_2(E)/B_0(E)) \delta E$.  
$B_2(E)/B_0(E)$ was given by the polynomial for $E$ and $E^2$, where the  coefficients were determined by the $\chi^2$ fit analysis.  

The gain fluctuation was typically within 40 keV.
The systematic uncertainty due to the gain fluctuation of the E detector was evaluated using the same procedure as that for the energy-scale uncertainty. 
 
For a pileup event caused by two $\beta$ rays, the obtained alignment correlation term is determined on the basis of the contributions of the two $\beta$ rays at their respective energies. This effect was evaluated as a systematic uncertainty by integrating its contribution over the energy of two $\beta$ rays.

\begin{table}[hbt]
\caption{\label{Syst9MeV}
Systematic uncertainties of the alignment correlation term at 9MeV. }
\begin{ruledtabular}
\begin{tabular}{ccc}
&$^8$Li &$^8$B \\
&$\times 10^{-2}$&$\times 10^{-2}$\\
\hline
Polarization relaxation &0.002&0.003\\
Third-order orientation &0.015&0.026\\
\begin{minipage}{5cm}
\vspace{2mm}
Uncertainty of $B_2/B_0$ 
in detector-response correction
\vspace{2mm}
\end{minipage}
&0.010&0.007\\
\begin{minipage}{5cm}
\vspace{2mm}
Uncertainty of matrix elements 
in $B_1/B_0$ correction
\vspace{2mm}
\end{minipage}
&0.107&0.133\\
Position of beam spot and catcher &0.011&0.027\\
Position of $\Delta$E detector and catcher &0.015&0.039\\
Low-energy tail of detector response &0.050&0.034 \\
Detector resolution &0.021&0.017 \\
Energy scaling & 0.053 & 0.001\\
Gain fluctuation & 0.028& $<0.001$\\
Pileup &0.001&$<0.001$\\
Background &$<0.001$&$<0.001$\\
\hline
Total &0.137&0.148
\end{tabular}
\end{ruledtabular}
\end{table}

\section{Results and discussion}

After the obtained alignment correlation terms are shown, 
the results given by Eq.~(\ref{Eq:diffsum}) are compared with the CVC prediction.  
And then the weighted mean value of end-point energy over final-state distribution is described,  
which is used when the matrix elements are determined from the alignment correlation terms and
the $\beta$-$\alpha$ correlation terms. 
Finally, the extraction of the weak magnetism and the second-forbidden term 
is described and these terms are compared with the CVC prediction. 

\subsection{Alignment correlation terms and $\beta$-$\alpha$ correlation terms}

\begin{figure}
\includegraphics[scale=0.9]{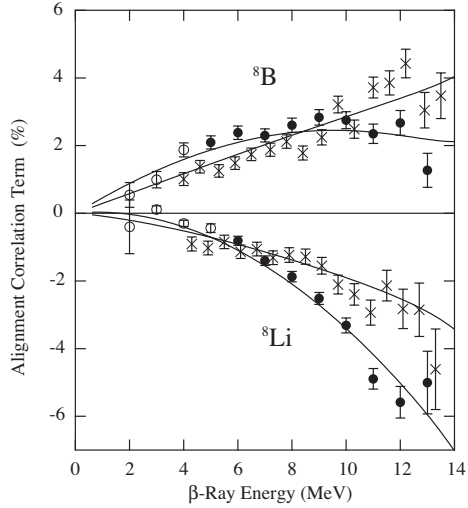}
\caption{Alignment correlation terms 
and $\beta$-$\alpha$ correlation terms. 
The circles are the alignment correlation terms and
the crosses are the $\beta$-$\alpha$ angular correlation terms. 
The open circles are not used for the derivation of the matrix elements. 
The lines are the best-fit curves.}
\label{fig:aE}
\end{figure}

The alignment correlation terms that were obtained are
shown in Fig.~\ref{fig:aE}. 
The statistical uncertainty of the alignment $\Delta{\cal A}_{1+2}$ 
in Eq.~(\ref{R1R2}) could shift all data points of the alignment correlation term 
in the same direction. 
The statistical uncertainty of $\Delta{\cal A}_{1+2}$  
is not included in each data point of Fig.~\ref{fig:aE} 
in order to retain a statistical fluctuation among the different points; 
however, the statistical uncertainties of the final results, such as $a^\beta_{\rm WM}$, 
include the statistical uncertainty of $\Delta{\cal A}_{1+2}$.  

\begin{figure} 
\includegraphics[scale=0.9]{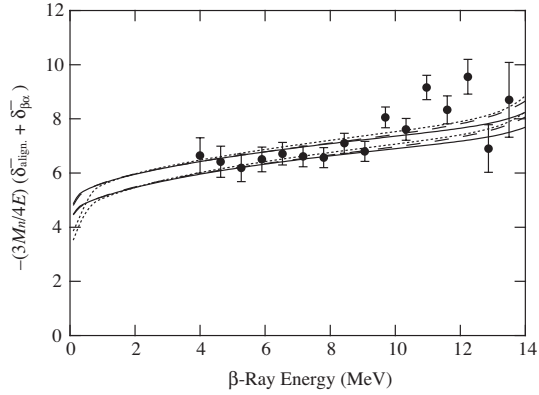}
\caption{\label{bAc}
Weak magnetism term derived from the $\beta$-ray correlation terms. The CVC 
predictions are shown by the $1\sigma$ error band.
The solid, dashed and dotted bands are the present, 
De Braeckeleer's \cite{DeBraeckeleer95}, and 
Winter's \cite{Winter03, Winter06} predictions, respectively. 
Winter's prediction was re-evaluated 
using the mirror-averaged end-point energy.
}
\end{figure}

The $\beta$-$\alpha$ correlation terms $-\frac{2}{3}p_\pm(E)$
\cite{McKeown80} are also shown as crosses. 
The weak magnetism term, 
$-(3M_n/4E)(\delta^-_{\rm align.} +\delta^-_{\beta\alpha})=b/Ac$, 
was derived combining the two types of correlation term 
shown in Fig.~\ref{bAc}. 
Figure \ref{bAc} also reflects a reanalysis using the same energy bin
as the $\beta$-$\alpha$ correlation terms \cite{McKeown80}.

\subsection{CVC prediction}
\label{CVCprediction}

The experimental $b/Ac$ results shown in Fig.~\ref{bAc} indicates a slight $E$ dependence. 
The CVC prediction of an energy dependent $b/Ac$ has in previous studies been indicated and 
described by introducing the dependence 
into the matrix elements, $b(E_x)$ and $c(E_x)$, 
of the final-state energy $E_x$ in $^8$Be 
\cite{DeBraeckeleer95,Nathan75,Paul75,Bowles78}. 
The final-state energy distributes widely because 
several states with spin and parity of 2$^+$ are mixed because of the wide decay width. 
This final-state distribution can be formulated using the $R$-matrix theory
with four final states \cite{Warburton86, Bhattacharya06}. 
We re-evaluated the CVC prediction in Ref.~\cite{Sumikama08-1} by using 
the analog-$\gamma$-decay measurement by De Braeckeleer 
{\it et al.}~\cite{DeBraeckeleer95}, and 
the recent measurement of 
the $\beta$-delayed-$\alpha$ energy spectra from $^8$Li and $^8$B
by Bhattacharya {\it et al.}~\cite{Bhattacharya06}.
The procedure for this re-evaluation was same 
as for the previous work~\cite{DeBraeckeleer95}
except for the number of final states; 
three final states were used in it, 
while four final states were used in the present evaluation, 
similar to that for the Gamow-Teller matrix element $c(E_x)$ 
in Ref.~\cite{Bhattacharya06}. 
The procedure is summarized below. 

The $E_x$ dependence of $c(E_x)$ 
gives the final-state distribution for the $\beta$ decay,
i.e., the delayed $\alpha$ energy spectrum.
The mirror-averaged $c(E_x)$ was determined 
from the delayed $\alpha$ energy spectra 
of $^8$Li and $^8$B based on the $R$-matrix formalism 
by Bhattacharya \textit{et al.}~\cite{Bhattacharya06}. 

$b(E_x)$ is given by the isovector M1 transition 
strength of the analog-$\gamma$ decay, based on the strong CVC.
The isobaric analog state in $^8$Be was produced 
using the $^4$He($\alpha$, $\gamma$) reaction  
and the de-excited $\gamma$ ray was measured \cite{DeBraeckeleer95}. 
The $E_x$ dependence of $b(E_x)$ gives
the final-state distribution in the analog-$\gamma$ decay, 
which has been measured through the $\gamma$-ray energy spectra 
shown in Fig.~4 of Ref.~\cite{DeBraeckeleer95}. 
The matrix elements, ${\cal M}_1^\gamma$ and $R_{\gamma}$ in $b(E_x)$, 
gives the $E_x$ dependence of $b(E_x)$ and 
were determined using three final states in Ref.~\cite{DeBraeckeleer95}. 
The ${\cal M}_1^\gamma$  is the weak magnetism matrix element 
for the transition to the first excited state,   
and $R_{\gamma}$ is the ratio, ${\cal M}_{16}^\gamma/{\cal M}_1^\gamma$, 
where ${\cal M}_{16}^\gamma$ represents 
the transition to an isospin doublet at 16 MeV.
These matrix elements were re-determined for the four final states
so as to reproduce the $\gamma$-ray energy spectra, 
which were ${\cal M}_1^\gamma=8.71\pm0.28$ and $R_\gamma=1.5\pm1.4$.

The $E$ dependent $b/Ac$ is given 
in \cite{DeBraeckeleer95} by the weighted average as 
\begin{equation}
\label{eq;WM}
\frac{b}{Ac}\rightarrow\frac{\int b(E_x)c(E_x)(E_{\max}-E_x-E)dE_x}
     {A\int c^2(E_x)(E_{\max}-E_x-E)dE_x}.
\end{equation}
Here, $c(E_x)$ and $E_{\rm max}$ averaged between the mirror pair
were used, 
because $b/Ac$ was derived from the $\beta$-ray angular correlations
of both $^8$Li and $^8$B. 
The CVC prediction of $b/Ac$ determined from Eq.~(\ref{eq;WM}) is 
shown in Fig.~\ref{bAc}.

The CVC prediction determined by De Braeckeleer {\it et al.}~\cite{DeBraeckeleer95}
and Winter {\it et al.}~\cite{Winter03, Winter06} are also shown in Fig.~\ref{bAc}. 
The previous predictions have a problem in regards to final-state treatment. 
The delayed-$\alpha$ spectra have been reproduced very well 
using four final states \cite{Warburton86, Bhattacharya06}. 
De Braeckeleer {\it et al.}, however, used three states 
for both $b(E_x)$ and $c(E_x)$, and  
Winter {\it et al.}~used three states only for $b(E_x)$.
The present CVC prediction was slightly smaller than 
the previous predictions at a higher energy region. 

The transition to the first excited state of $^8$Be was predominant 
for the analog-$\gamma$ transition \cite{DeBraeckeleer95}. 
Therefore, 
for a comparison between the $b/Ac$ extracted from the $\beta$ decay 
and its CVC prediction, 
the weak magnetism for the first excited state, $a_{\rm WM}$, was used. 
This was given 
by the matrix elements of the transition to the first excited state,  
$a_{\rm WM}^{\beta}={\cal M}^\beta_1/A{\cal M}_{\rm GT1}^\beta$
for $\beta$ decay and 
$a_{\rm WM}^{\rm CVC}={\cal M}^\gamma_1/A{\cal M}_{\rm GT1}^\beta$
for the CVC prediction, 
where ${\cal M}^\beta_1$ and  
${\cal M}_{\rm GT1}^\beta$ are 
the weak magnetism, $b$, and Gamow-Teller, $c$, matrix elements
for the weak transition to the first excited state, respectively. 
The expression for $b/Ac$ using ${\cal M}_1$ and  ${\cal M}_{\rm GT1}$
was given in Ref.~\cite{DeBraeckeleer95}. 
The CVC prediction was determined to be 
$a_{\rm WM}^{\rm CVC}=7.3\pm0.2$ based on
${\cal M}^\gamma_1=8.71\pm 0.28$ 
and  the mirror-averaged ${\cal M}^\beta_{\rm GT1}=0.1496\pm0.0005$ \cite{Bhattacharya06}.
The CVC prediction of $f$ was determined 
by the isovector M1/E2 ratio $\delta_1=0.01\pm 0.03$ \cite{DeBraeckeleer95}
as $a_{\rm WE2}^{\rm CVC}=\sqrt{10/3}\delta_1 a_{\rm WM}^{\rm CVC}=0.1\pm0.4$. 
The values are summarized in Tables \ref{ResultsGamma} and \ref{Results}.

\begin{table}[bth]
\caption{\label{ResultsGamma} Decay widths and matrix elements for the $\gamma$ decay from the isobaric analog state in $^{8}$Be. 
$\Gamma_{M1}^{T=1}$ is the decay width for the isovector component of the M1 transition from the isobaric analog state ($T=1$). $\delta_1$ is the isovector M1/E2 ratio from the isobaric analog state. 
Definition of ${\cal M}^{\gamma}_1$ and $R_1^\gamma$ is described in the text. Average value ${\cal M}^{\beta}_{\rm GT1}$ of the Gamow-Teller matrix elements  of $^8$Li and $^8$B is also shown.  
}

\begin{ruledtabular}
\begin{tabular}{lclclc}
Analog $\gamma$ decay & Value & Matrix Element & Value\\
\hline
$\Gamma_{M1}^{T=1}$ \cite{DeBraeckeleer95} & $2.80\pm0.18$ eV &
${\cal M}^\gamma_1$ \cite{DeBraeckeleer95}\footnotemark[1]
& $8.7\pm0.3$\\
$\Gamma_{M1}^{T=1}$ \cite{Bowles78}\footnotemark[2] & $3.6\pm0.3$ eV &
${\cal M}^\gamma_1$ \cite{Bowles78}\footnotemark[3]
& $9.9\pm0.6$\\
$\Gamma_{M1}^{T=1}$ \cite{Paul75}\footnotemark[2] & $4.1\pm0.6$ eV &
${\cal M}^\gamma_1$ \cite{Paul75}\footnotemark[3]
& $10.5\pm0.9$\\
$\delta_1$ \cite{DeBraeckeleer95} &$0.01\pm0.03$&
$R^\gamma_1$ \cite{DeBraeckeleer95}\footnotemark[1] & $1.5\pm1.4$\\
$\delta_1$ \cite{Bowles78}\footnotemark[2]  
 &$0.14\pm0.03$&
${\cal M}^\beta_{\rm GT1}$ \cite{Bhattacharya06}
& $0.1496\ \ \ $\\
 &&& $\ \pm 0.0005$\\
\end{tabular}
\end{ruledtabular}
\footnotetext[1]{Reanalyzed in the present work using the four final states in the $R$-matrix formalism. }
\footnotetext[2]{Reanalyzed in Ref.~\cite{DeBraeckeleer95}. }
\footnotetext[3]{Calculated from ${\cal M}^\gamma_1$ of Refs.~\cite{DeBraeckeleer95} and $\Gamma_{M1}^{T=1}$ of Ref.~\cite{DeBraeckeleer95, Bowles78} or Refs.~\cite{DeBraeckeleer95, Paul75} using the relation ${\cal M}^\gamma_1 \propto \sqrt{\Gamma_{M1}^{T=1}}$. }
\end{table}

\subsection{End-point energy}
\begin{figure}
\includegraphics[scale=1.]{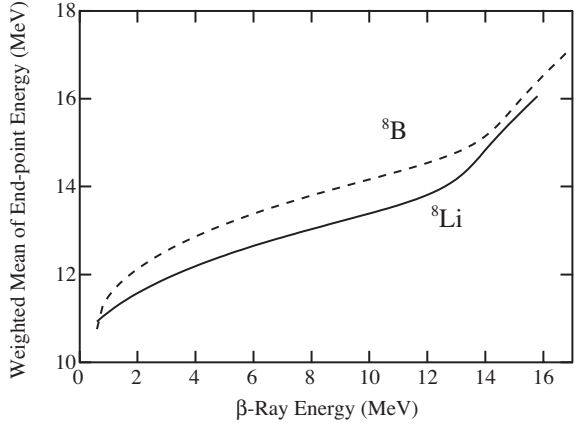}
\caption{Weighted mean value of the end-point energy over the broad-final-state distribution as a function of $\beta$-ray energy.}
\label{fig:endpoint}
\end{figure}

The end-point energy of the $\beta$ ray is not a constant because of the broad final state. 
The alignment correlation terms  and the $\beta$-$\alpha$ correlation terms 
given in Eq.~(\ref{Eq:B2B0}) were measured as a function of $\beta$-ray energy 
without measurement of the end-point energy.  
Therefore, the end-point energy was averaged over the final-state-energy distribution. 
The weight is the product of $pE(E_0(E_x) - E)^2$ and the final-state distribution $ c^2(E_x)$ of the $\beta$ decay. 
When a certain $\beta$-ray energy is chosen,  the weighted mean value of the end-point energy is given by   
\begin{eqnarray}
\overline{E}_0(E) &=& \frac{\int pE(E_0(E_x) -E)^2 c^2(E_x) E_0(E_x) dE_x}
{\int pE(E_0(E_x) -E)^2 c^2(E_x) dE_x} \nonumber \\
              &=& \frac{\int (E_0(E_x) -E)^2 c^2(E_x) E_0( E_x) dE_x}
{\int (E_0( E_x) -E)^2 c^2(E_x) dE_x},
\end{eqnarray}
where $E_0( E_x) = E_{\rm max} - E_x$ and the integral range is from 0 to ($E_{\rm max} - E)$. 
Figure \ref{fig:endpoint} shows $\overline{E}_0(E)$ calculated using $c(E_x)$ determined in Ref.~\cite{Bhattacharya06}. This $\overline{E}_0(E)$  was used in the analysis to determine the matrix elements. 

\subsection{Weak magnetism and second-forbidden terms from the weak vector current}
\label{extraction}

The mirror difference $\delta^-$ consists of
$b/Ac$ and a small contribution of $j_2/A^2c$ 
due to the mirror asymmetry of $E_0$.
To avoid the influence of this mirror asymmetry, 
the $\chi^2$ fit analysis was performed 
simultaneously on the four correlation terms, i.e., 
both alignment correlation terms and 
$\beta$-$\alpha$ angular correlation terms of $^8$Li and $^8$B.
The $E_x$ dependent $b/Ac$ in the $\beta$-ray angular correlation terms
was given by the same formula as the CVC prediction of $b/Ac$.  
$a_{\rm WM}^\beta=
{\cal M}^\beta_1 /A{\cal M}_{\rm GT1}^\beta$ was used 
as a free parameter for the $\chi^2$ fit analysis, where
$R_\beta$ was assumed to be the same as $R_\gamma$.
The $E$ dependences of $a_{\rm WE2}^\beta$, $d_{\rm I}/Ac$, 
$j_2/A^2c$, and $j_3/A^2c$ were not clearly seen 
in the $\beta$-ray correlation terms 
because of the relatively large statistical uncertainties. 
These terms were assumed to be constant  
and were chosen as free parameters for the $\chi^2$ fit analysis. 
The obtained terms were 
considered as the value averaged over the analyzed energy region. 
The best-fit curves are shown in Fig.~\ref{fig:aE} and
the results are summarized in Table \ref{Results}. 
The weak magnetism and the second-forbidden terms 
were $a_{\rm WM}^\beta=7.54\pm0.12$(stat.)$\pm0.15$(syst.) and 
$a_{\rm WE2}^\beta=1.0\pm0.2$(stat.)$\pm0.2$(syst), respectively. 
The systematic uncertainty because of the $E$ dependence of $a_{\rm WE2}^\beta$
was estimated to be 0.05 for $a_{\rm WE2}^\beta$ 
by assuming that the $E_x$ dependence of $f(E_x)$ was the same as $b(E_x)$.  
The other systematic uncertainties in the alignment correlation terms 
and the $\beta$-$\alpha$ correlation terms were independently propagated
to those in $a_{\rm WM}^\beta$ and $a_{\rm WE2}^\beta$ by performing the $\chi^2$ fit analysis
for the data applied to the different correction factors.  
$a^\beta_{\rm WM}$ was consistent with
the CVC prediction from De Braeckeleer's data, i.e., 
$a^\beta_{\rm WM}/a^{\rm CVC}_{\rm WM} = 1.03\pm0.04$. 
However, the present $a_{\rm WE2}^\beta$ is inconsistent with 
the De Braeckeleer's data, $a_{\rm WE2}^{\rm CVC}=0.1\pm0.4$. 
The deviation of $a_{\rm WE2}$ was $1.8\sigma$ 
as $a_{\rm WE2}^\beta-a_{\rm WE2}^{\rm CVC}=0.9\pm0.5$.

\begin{table}[tbh]
\caption{\label{Results}Ratio of matrix elements contributing to the $\beta$-ray 
angular correlations. 
The CVC predictions are also shown. 
$a_{\rm WM}^{\rm CVC}={\cal M}_1^\gamma/A {\cal M}_{\rm GT1}^\beta$ and
$a_{\rm WE2}^{\rm CVC}=\sqrt{10/3} \delta_1 a_{\rm WM}^{\rm CVC}$. 
}
\begin{ruledtabular}
\begin{tabular}{lclc}
Matrix Element & Value &Matrix Element & Value\\
\hline
$a_{\rm WM}^\beta$ & $7.5\pm0.2$  &
$a_{\rm WM}^{\rm CVC}$ \cite{DeBraeckeleer95}
& $7.3\pm0.2$\\
$a_{\rm WE2}^\beta$  & $1.0\pm0.3$&
$a_{\rm WM}^{\rm CVC}$ \cite{Bowles78}  
& $8.3\pm0.5$\\
$d_{\rm I}/Ac$     & $5.5\pm1.7$  &
$a_{\rm WM}^{\rm CVC}$ \cite{Paul75}
& $8.8\pm0.7$\\
$j_2/A^2c$         & $-490\pm70$  &
$a_{\rm WE2}^{\rm CVC}$ \cite{DeBraeckeleer95}   
& $0.1\pm0.4$\\
$j_3/A^2c$         & $-980\pm280$ &
$a_{\rm WE2}^{\rm CVC}$ \cite{Bowles78}     
& $2.1\pm0.5$\\
\end{tabular}
\end{ruledtabular}
\end{table}

We compared these results with the other analog-$\gamma$-decay measurements 
by Bowles and Garvey \cite{Bowles78} and Paul {\it et al.}~\cite{Paul75}.  
The CVC prediction was $a_{\rm WM}^{\rm CVC}=8.3\pm0.5$
and  $a_{\rm WE2}^{\rm CVC}=2.1\pm0.5$ 
for Bowles' data, and 
$a_{\rm WM}^{\rm CVC}=8.8\pm0.7$
for Paul's data. 
These predictions were inconsistent with the De Braeckeleer's data
and also with the present $\beta$-decay results; that is, 
both $a_{\rm WM}^{\rm CVC}$ were larger than $a_{\rm WM}^{\beta}$, 
and the deviation of $a_{\rm WE2}$ was  $1.8\sigma$, as $-1.1\pm0.6$.  
It was pointed out 
by De Braeckeleer {\it et al.}~\cite{DeBraeckeleer95}
that there were problems in these measurements 
in regards to the absolute cross section, the photon angular distribution 
and the neutron background. 
The difference between the two $a^{\rm CVC}_{\rm WE2}$ 
was due to deviation of $\delta_1$, i.e., 
$0.01\pm0.03$ \cite{DeBraeckeleer95} compared to
$0.14\pm0.03$ \cite{Bowles78,DeBraeckeleer95}.  
This deviation was determined via the relatively difficult measurement of 
the photon angular distribution. 
The inconsistency might be due to an underestimated background for 
the photon angular distribution. 
Although the CVC prediction by De Braeckeleer {\it et al.}~was 
adopted in the present work, 
De Braeckeleer's data need to be confirmed with 
more accurate measurements.  

\section{Summary}
The nuclear-spin-aligned nuclei $^8$Li and $^8$B were produced 
from spin-polarized nuclei using the $\beta$-NMR technique
to test the strong CVC  at a zero momentum transfer limit. 
The strong CVC could be tested 
for the second-forbidden transition for the first time. 
The alignment correlation terms for the $\beta$-ray angular distribution 
were determined using 
both positively and negatively aligned nuclei. 
The weak magnetism and the second-forbidden terms originating
from the weak vector current were determined 
by combining the present alignment correlation terms and 
the previously known $\beta$-$\alpha$ angular correlation terms. 
The CVC predictions of the weak magnetism and the 
second-forbidden terms were re-evaluated using the most precise data set
of the analog-$\gamma$ decay in $^8$Be.  
Although the weak magnetism term was consistent 
with the CVC prediction obtained from the isovector-M1-transition strength, 
the second-forbidden term was inconsistent with 
that from the isovector-E2-transition strength.
For more reliable tests for the second-forbidden transition, 
the CVC predictions need to be confirmed 
by more accurate measurements 
especially with regard to the isovector M1/E2 ratio $\delta_1$.

\begin{acknowledgments}
This work was supported by KAKENHI  (21740189). 
\end{acknowledgments}

%


\end{document}